# Infrared spectra of $C_2H_4$ dimer and trimer

**A.J. Barclay, [a] K. Esteki,[a] A.R.W. McKellar, [b] and N. Moazzen-Ahmadi [a,*]**

[a] Department of Physics and Astronomy, University of Calgary, 2500 University Drive
North West, Calgary, Alberta T2N 1N4, Canada

[b] National Research Council of Canada, Ottawa, Ontario K1A 0R6, Canada

Suggested running head:   $C_2H_4$ dimer and trimer

Address for correspondence:      Dr. N. Moazzen-Ahmadi
                                 Department of Physics and Astronomy,
                                 University of Calgary,
                                 2500 University Drive North West,
                                 Calgary, Alberta T2N 1N4,
                                 Canada

* Corresponding author. Fax: +1-403-289-3331; e-mail: ahmadi@phas.ucalgary.ca (N.

Moazzen-Ahmadi)

**Abstract**

Spectra of ethylene dimers and trimers are studied in the $\nu_{11}$ and (for the dimer) $\nu_9$ fundamental band regions of $C_2H_4$ ($\approx$2990 and 3100 cm$^{-1}$) using a tunable optical parametric oscillator source to probe a pulsed supersonic slit jet expansion. The deuterated trimer has been observed previously, but this represents the first rotationally resolved spectrum of $(C_2H_4)_3$. The results support the previously determined cross-shaped ($D_{2d}$) dimer and barrel-shaped ($C_{3h}$ or $C_3$) trimer structures. However, the dimer spectrum in the $\nu_9$ fundamental region of $C_2H_4$ is apparently very perturbed and a previous rotational analysis is not well verified.

## 1. Introduction

Ethylene clusters are the simplest π-π organic systems that can serve as test cases for nonbonding interaction theories. Ethylene dimer and larger ethylene clusters are also of interest for astrophysical applications. These complexes may exist in the atmospheres of giant planets and their satellites [1,2], where they could contribute to the IR absorption spectra.

There have been two previous analyses of ethylene dimer spectra with rotational line assignments. In 1995, Chan et al. [3] studied $(C_2H_4)_2$ in the regions of the $C_2H_4$ $\nu_9$ (≈3100 cm$^{-1}$) and $\nu_{11}$ (≈2990 cm$^{-1}$) fundamental bands, and in 2012, our group studied $(C_2D_4)_2$ in the region of the $C_2D_4$ $\nu_{11}$ (≈2200 cm$^{-1}$) band [4]. The latter paper also included the first spectroscopic study of an ethylene trimer, $(C_2D_4)_3$. Even prior to 1995 there had been an extensive history of ethylene cluster research, sometimes marked by controversy about vibrational lifetimes [5-13]. Chan et al. established that the dimer is a symmetric top, with an edge-on cross-shaped structure having $D_{2d}$ symmetry and four equivalent hydrogen bonds. This geometry is supported by *ab initio* calculations [3,14-17] and matrix isolation spectra [18]. For the trimer, we also observed a symmetric top spectrum and proposed a structure with three equivalent ethylene monomers in a barrel (or ring) configuration with $C_{3h}$ or $C_3$ symmetry. The dimer and trimer are illustrated here in Fig. 1.

In the present paper, we detect the spectrum of the normal trimer, $(C_2H_4)_3$, in the $\nu_{11}$ region and also re-examine the spectrum of $(C_2H_4)_2$. Our results are entirely consistent with the previously proposed [3, 4] trimer and dimer structures, while providing the first molecular constants for $(C_2H_4)_3$ and improved parameters for $(C_2H_4)_2$. Both spectra exhibit homogeneous broadening indicative of finite upper state lifetimes. For the dimer, the line widths appear to vary from about 0.015 to 0.030 cm$^{-1}$ (Chan et al. reported 0.030 cm$^{-1}$). For the trimer, widths are about 0.025 cm$^{-1}$. These widths are considerably larger than those we observed previously [4] for the fully deuterated species, which were about 0.006 cm$^{-1}$ for the dimer and 0.005 cm$^{-1}$ for the trimer. We find that the dimer spectrum in the $\nu_9$ region appears to be highly perturbed, and our observation is not consistent with the rotational analysis given in [3].

## 2. Results

The spectra were recorded using our previously described pulsed supersonic slit jet apparatus [19-21] together with a Lockheed Martin Aculight Argos tunable optical parametric oscillator source. The expansion gas was a dilute mixture (0.2 – 0.4%) of $C_2H_4$ in helium with a backing pressure of about 10 atm. Simultaneous etalon and room temperature OCS reference gas spectra were recorded to perform frequency interpolation and calibration. The PGOPHER computer program [22] was used for simulation and fitting of the spectra.

### 2.1. *Ethylene dimer*

The observed ethylene dimer spectrum in the $\nu_{11}$ region is illustrated in the top trace of Fig. 2. The breaks in this spectrum correspond to the positions of sharp ethylene monomer absorption lines. Even at the low effective rotational temperature ($\approx$2.5 K) of our supersonic jet, there are many such monomer lines, arising not only from the $\nu_{11}$ fundamental band itself but also from various combination and hot bands which have been assigned to a greater or lesser degree [23-26]. Since our effective temperature is higher than that of [3] (2.5 vs. 0.9 K), our spectrum is more extended. Thus, for example, we clearly observe previously absent $Q$-branches with $K = 3 \leftarrow 2$ and $1 \leftarrow 2$ (see Fig. 2). It happens that $A \approx 7B$ for $(C_2H_4)_2$, and the manifestation of this "accidental" coincidence is that there are five $P$- or $R$-branch lines between two successive $Q$-branch features as in Fig. 2. A similar effect occurs for $(C_2D_4)_2$, where $A \approx 6B$ and there are four lines between two successive $Q$-branch features [4].

The dimer spectrum is a doubly-degenerate symmetric top perpendicular ($\Delta K = \pm 1$)

band, since the ethylene $\nu_{11}$ transition dipole is parallel to the monomer C-C axis and thus perpendicular to the dimer symmetry axis. The parameters resulting from a fit to the spectrum are listed in Table 1, and the corresponding simulated spectrum is shown in Fig. 2. Observed line positions are given as Supplementary Data. The simulation includes the nuclear spin weights (76:60:120 for $A_1$, $B_1$:$A_2$, $B_2$:$E$ levels), but as noted in [3] this detail has little visible effect on the spectrum. Since some peaks in the spectrum are assigned to many unresolved transitions, the statistical uncertainties in Table 1 probably underestimate the true error limits. Moreover, it is important to note that it is not possible to independently determine the Coriolis parameter $\zeta$ and the $A$ (or $C$) constant of a symmetric rotor from a normal infrared spectrum. Our fit to the present band assumes zero Coriolis interaction in the degenerate excited state, as was also assumed in [3] and [4]. This is usually considered a reasonable approximation for a weakly-bound van der Waals molecule [27], but it still should be kept in mind that the $A$-values in Table 1 really include the unknown Coriolis effects. Having said that, we note that the fitted value for $A''$ (= 0.4990 cm$^{-1}$) of the dimer is indeed very close to half the value of $B''$ for the monomer (= 0.5005 cm$^{-1}$) which is the expected value based on a rigid $D_{2d}$ structure (Fig. 1). The difference (0.3 %) is very similar to that observed for $(C_2D_4)_2$ [4].

The observed $B''$ constant yields a value of 3.888 Å for the separation of the centers of mass of the monomers in $(C_2H_4)_2$, assuming the monomer geometry remains unchanged in the dimer, whereas Chan et al. [3] derived a value of 3.922 Å [3] from their $B$-value. For comparison, the separation we determined [4] for $(C_2D_4)_2$ was 3.867 Å. The larger value for $(C_2H_4)_2$ can be explained by the different zero-point wavefunctions of the two isotopologues in the anharmonic intermolecular potential.

Chan et al. [3] also observed an ethylene dimer spectrum at 3102 cm$^{-1}$ in the $C_2H_4$ $\nu_9$ region. Our spectrum is shown in Fig. 3, where it must be admitted that the detail and clarity of our result is not good (nor was that of [3])! A parallel band ($\Delta K = 0$) is expected here, since the ethylene $\nu_{11}$ transition dipole is perpendicular to the monomer C-C axis and thus parallel to the dimer symmetry axis. Chan et al. [3] obtained a moderately good parallel band simulation of their 0.9 K spectrum by assuming a rather large change in $A$-value for the upper state ($A'' = 0.4965$, $A' = 0.465$ cm$^{-1}$). However, Fig. 3 illustrates that their parameters (blue trace) do not work at all well for simulating our 2.5 K spectrum. Also shown in the figure is a nominal parallel band simulation (red trace) with no change in rotational parameters between ground and excited states. This nominal simulation bears some resemblance to the observed spectrum, but there are many differences! We do not have a good explanation for this band, and can only suppose that the upper state is highly perturbed. Such perturbations are certainly possible, since there are many ethylene vibrational states in this region, including for example $\nu_5$ ($\approx 22$ cm$^{-1}$ below $\nu_9$), $2\nu_{10}+\nu_{12}$ (only 0.55 cm$^{-1}$ below), and $\nu_3+\nu_8+\nu_{10}$ (4.4 cm$^{-1}$ above) [28].

2.2. *Ethylene trimer*

Another band which was observed in a region just below the $\nu_{11}$ dimer band is illustrated in Fig. 4, and we assign this to the ethylene trimer. This $(C_2H_4)_3$ band was not observed by Chan et al. [3], but it is very much analogous to the $(C_2D_4)_3$ band observed in [4]. As discussed previously [4], we think the trimer has a barrel shaped structure similar to that shown in Fig. 1, which is a symmetric top with equivalent monomers and a $C_3$ rotational axis. The present result for $(C_2H_4)_3$ is entirely consistent, and thus supports the previous analysis [4]. But unfortunately, the spectra are not sensitive to the exact details of the structure, specifically the twist angle of the monomer planes relative to the line connecting the monomer and trimer centers of mass. It is also possible that the monomers could twist so that their C-C axes deviate from being parallel to the trimer $C_3$ symmetry axis [29].

But this deviation, if any, is small so the trimer is observed to have a parallel band in the $\nu_{11}$ region, opposite to the situation for the dimer. This means that the trimer $C$-value is not determinable from the spectrum, except very approximately from the intensity of the $Q$-branch relative to the $P$- and $R$-branches. Thus in our analysis, we used a calculated $C$-value consistent with that used previously for $(C_2D_4)_3$ [4], as described in the following paragraph. The resulting parameters for $(C_2H_4)_3$ are listed in Table 2, and the observed line positions are given as Supplementary Data. The simulation in Fig. 4 uses appropriate spin weights (1736:2720 for $A$:$E$ levels), but as with the dimer this does not significantly affect the observed spectrum. We have not detected a spectrum attributable to ethylene trimer in the $\nu_9$ region.

The nominal ethylene trimer structure from Ref. [4] has $C_{3h}$ symmetry with parallel monomer axes, a 30° twist angle, and an intermolecular separation of 3.947 Å, and it was this structure that determined the value of $C''$ (= 0.03770 cm$^{-1}$) used here. The predicted value of $B''$ for $(C_2H_4)_3$ from this structure is 0.06150 cm$^{-1}$, as compared to our fitted value of 0.06082 cm$^{-1}$ (Table 2). Conversely, the fitted $B''$-value implies a slightly larger intermolecular separation of 3.974 Å for $(C_2H_4)_3$. To summarize, we conclude that the assumed $C_{3h}$ trimer structure predicts the $(C_2D_4)_3$ to $(C_2H_4)_3$ isotope shift in $B''$ pretty well, but with the effective intermolecular bond being slightly longer for $(C_2H_4)_3$. This of course is expected from the larger anharmonicity and zero-point motion in $(C_2H_4)_3$. If the monomer C-C axes are not parallel to the trimer symmetry axis, the experimental intermolecular distances would become slightly (≈0.03 Å) larger [4].

The arrows in Fig. 4 mark three unassigned features in the region of the trimer spectrum at about 2983.89, 2984.34, and 2984.89 cm$^{-1}$ which have widths roughly similar to those of the trimer (perhaps somewhat broader). There is some evidence (from gas mixture concentration dependence) that the 2984.34 cm$^{-1}$ feature might be due to a larger cluster (e.g. tetramer) but this is not conclusive. Some analogous unassigned features were also observed for $(C_2D_4)_3$ [4], where it was noted that they could possibly be due to $Q$-branches of a trimer hot band progression. Here we mention another possibility: assuming that the main trimer band arises from the in-phase $\nu_{11}$ vibration of the three monomers, then there could be another weaker trimer band for which two monomers are in-phase and the other out-of-phase.

### 3. Discussion and conclusions

Observed vibrational shifts and linewidths for ethylene dimers and trimers are summarized in Table 3. The $\nu_{11}$ dimer and trimer shifts for $C_2H_4$ are 1.29 and 1.42 times those for $C_2D_4$, roughly similar to the ratio of the vibrational frequencies themselves, 1.36. The $\nu_9$ dimer shift is about double that of $\nu_{11}$, but it should be regarded with caution due to the uncertainties of that band (Fig. 3). The linewidths in Table 3 are all significantly larger than the experimental resolution, indicating that they are due to vibrational predissociation (i.e. finite excited state lifetimes). Precise values for the widths are difficult to determine, in part because they seem to vary for different transitions in the same band. It is evident that dimer and trimer widths are roughly similar to each other, while those for $C_2H_4$ clusters are about 3 to 5 times larger than those for $C_2D_4$. Again, the $\nu_9$ dimer value is highly uncertain.

In conclusion, spectra of the ethylene dimer and trimer have been studied in the $C_2H_4$ $\nu_{11}$ (2990 cm$^{-1}$) and (for the dimer) $\nu_9$ (3100 cm$^{-1}$) regions. The results extend previous studies of $C_2H_4$ dimer [3] and $C_2D_4$ dimer and trimer [4] and confirm the previous conclusions, except that the $\nu_9$ dimer spectrum cannot be easily explained and is likely very perturbed. The dimer has a cross-shaped $D_{2d}$ structure with an effective intermolecular distance of 3.89 Å for $(C_2H_4)_2$. The trimer has a barrel-shaped structure with $C_{3h}$ (or $C_3$) symmetry and an effective intermolecular distance of 3.97 Å for $(C_2H_4)_3$.

### Acknowledgments

We gratefully acknowledge the financial support of the Natural Sciences and Engineering Research Council of Canada.

Table 1. Molecular parameters for $C_2H_4$ dimer in the $\nu_9$ region (in cm$^{-1}$). [a]

|  | Present work | Chan et al. [3] |
|---|---|---|
| $\nu_0$ | 2987.3149 (6) | 2987.310 |
| $A'$ | 0.49870 (10) | 0.4955 |
| $B'$ | 0.071600 (92) | 0.0700 |
| $10^6 \times D_J'$ | 7.3 (10) | — |
| $A''$ | 0.49898 (18) | 0.4965 |
| $B''$ | 0.071537 (90) | 0.0704 |
| $10^6 \times D_J''$ | 5.1 (11) | — |

[a]Statistical uncertainties in parentheses represent $1\sigma$ in units of the last quoted digit.

Table 2. Molecular parameters for $C_2H_4$ trimer in the $\nu_9$ region (in $cm^{-1}$). [a]

|  | Present work |
|---|---|
| $\nu_0$ | 2984.5649 (3) |
| $B'$ | 0.060824 (15) |
| $(C' - C'')$ | -0.000003 (10) |
| $10^6 \times D_J'$ | 1.05 (9) |
| $B''$ | 0.060821 (15) |
| $C''$ | 0.03770 [b] |
| $10^6 \times D_J''$ | 1.03 (9) |

[a]Statistical uncertainties in parentheses represent $1\sigma$ in units of

the last quoted digit.

[b] The parameter $C''$ was fixed at this calculated value (see text).

Table 3. Vibrational shifts and line widths for $C_2H_4$ and $C_2D_4$ dimer and trimer (in cm$^{-1}$). [a]

|  | $C_2H_4$ $\nu_9$ | $C_2H_4$ $\nu_{11}$ | $C_2D_4$ $\nu_{11}$ |
|---|---|---|---|
| Shift, dimer | -2.60 | -1.330 | -1.030 |
| Shift, trimer | — | -4.080 | -2.872 |
| Width, dimer | 0.04 | >0.020 | 0.006 - 0.010 |
| Width, trimer | — | >0.025 | >0.005 |

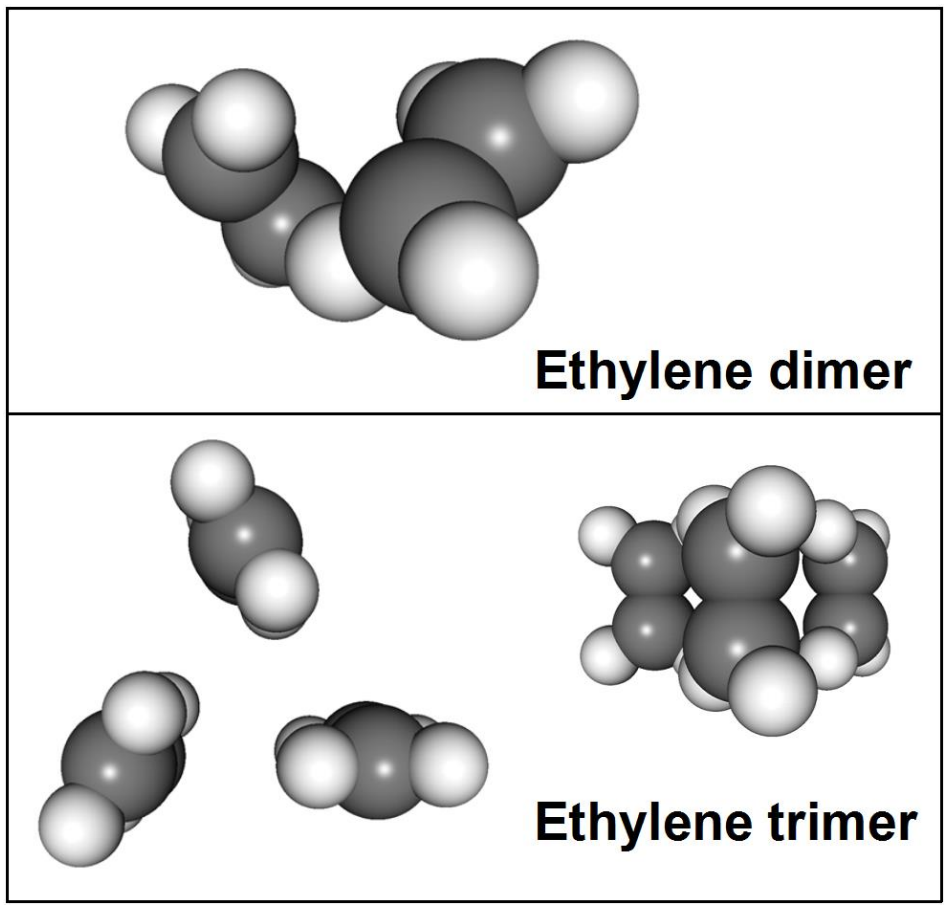

Figure 1: Ethylene dimer and trimer structures. The dimer is a symmetric top with $D_{2d}$ symmetry. The trimer, viewed from the "top" and "side", is a barrel shaped symmetric top with $C_{3h}$ symmetry as shown, or possibly $C_3$ symmetry if the monomer C-C axes are non-parallel (see Ref. [4]).

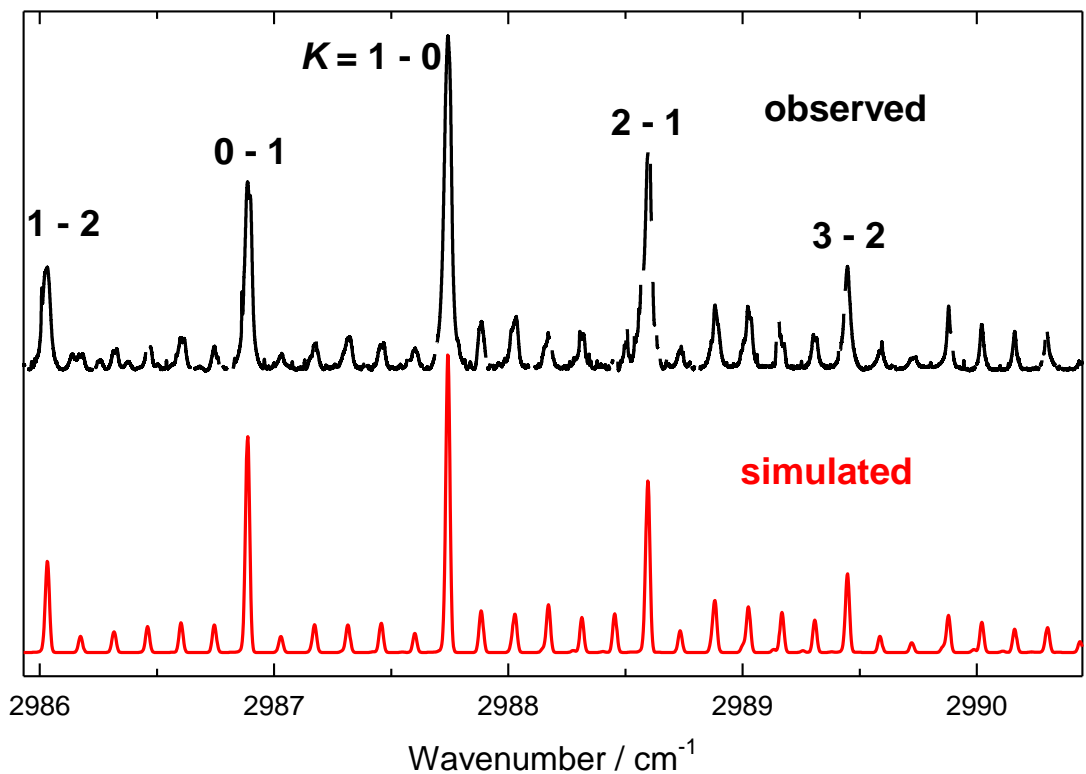

Figure 2: Observed and simulated spectra of ethylene dimer, $(C_2H_4)_2$, in the $C_2H_4$ $\nu_{11}$ region. Breaks in the observed trace correspond to sharp $C_2H_4$ monomer lines. The stronger dimer lines are $Q$-branches, with $K = 1 \leftarrow 2$, $0 \leftarrow 1$, $1 \leftarrow 0$, $2 \leftarrow 1$, and $3 \leftarrow 2$, as illustrated.

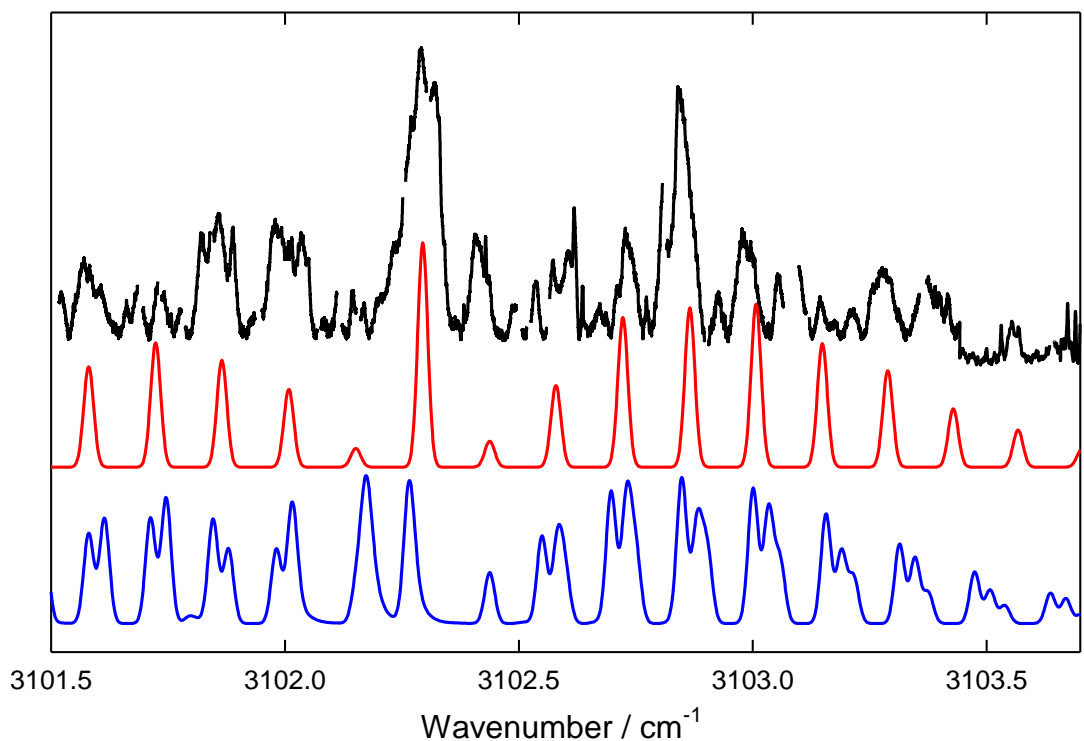

Figure 3: Observed and simulated ethylene dimer spectra in the $C_2H_4$ $\nu_9$ region. Breaks in the observed spectrum (top trace) correspond to sharp $C_2H_4$ monomer lines. The middle simulated trace (red) shows the "expected" parallel band, with upper and lower state parameters fixed to our ground state values (Table 1). The lower simulated trace (blue) uses the parameters of Chan et al. [3], which feature a large decrease in $A$-value (-0.032 cm$^{-1}$) between ground and excited states. Neither simulation accounts well for the observed spectrum (see text). For clarity, the broadening in both simulations (0.025 cm$^{-1}$ FWHM Gaussian) is deliberately made somewhat smaller than that present in the experimental spectrum.

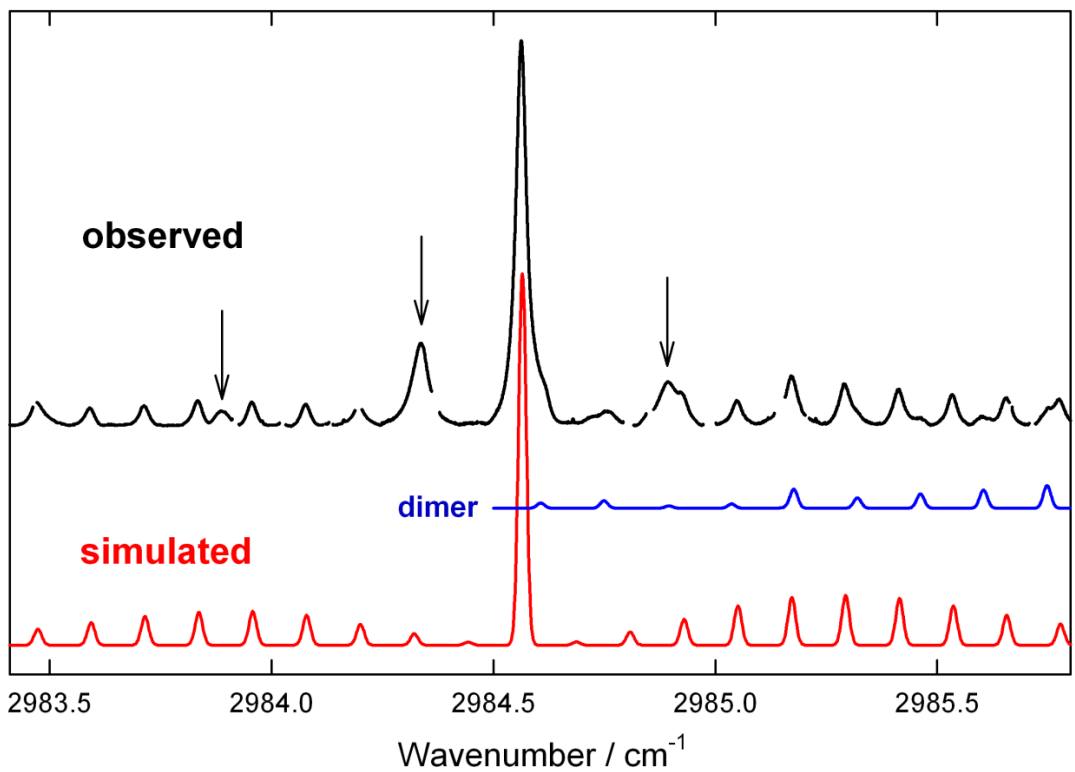

Figure 4: Observed and simulated spectra of ethylene trimer, $(C_2H_4)_3$. Breaks in the observed trace correspond to sharp $C_2H_4$ monomer lines. The arrows mark features which are probably due to other unassigned ethylene cluster bands.

Table A1. Observed transitions of C2H4 dimer (in units of 1/cm)

*Asterisks indicate the strong Q-branch features

In general, each line has a number of contributing transitions. The J, K assignments indicate the strongest contribution to each line.

**********************************

| J' | K' | J" | K" | Position |
|----|----|----|----|----------|

**********************************

| J' | K' | J" | K" | Position |
|----|----|----|----|----------|
| 3 | 1 | 3 | 2 | 2986.032* |
| 3 | 0 | 4 | 1 | 2986.315 |
| 2 | 0 | 3 | 1 | 2986.464 |
| 1 | 0 | 2 | 1 | 2986.602 |
| 0 | 0 | 1 | 1 | 2986.744 |
| 3 | 0 | 3 | 1 | 2986.887* |
| 4 | 1 | 5 | 0 | 2987.031 |
| 3 | 1 | 4 | 0 | 2987.172 |
| 3 | 0 | 2 | 1 | 2987.315 |
| 4 | 0 | 3 | 1 | 2987.455 |
| 5 | 0 | 4 | 1 | 2987.601 |
| 4 | 1 | 4 | 0 | 2987.743* |
| 1 | 1 | 0 | 0 | 2987.884 |
| 2 | 1 | 1 | 0 | 2988.031 |
| 7 | 1 | 6 | 0 | 2988.736 |
| 2 | 2 | 1 | 1 | 2988.884 |
| 3 | 2 | 2 | 1 | 2989.023 |
| 5 | 2 | 4 | 1 | 2989.312 |
| 6 | 2 | 5 | 1 | 2989.447* |
| 7 | 2 | 6 | 1 | 2989.589 |
| 8 | 2 | 7 | 1 | 2989.720 |
| 3 | 3 | 2 | 2 | 2989.880 |
| 4 | 3 | 3 | 2 | 2990.021 |
| 5 | 3 | 4 | 2 | 2990.161 |
| 7 | 3 | 6 | 2 | 2990.438 |
| 4 | 4 | 3 | 3 | 2990.877 |
| 6 | 4 | 5 | 3 | 2991.153 |
| 7 | 4 | 6 | 3 | 2991.292 |

**********************************

Table A2. Observed transitions of C2H4 trimer (in units of 1/cm)

******************

P(11)        2983.233

P(10)        2983.351

P( 9)        2983.476

P( 8)        2983.593

P( 7)        2983.715

P( 6)        2983.836

P( 5)        2983.957

P( 4)        2984.079

P( 3)        2984.198

Q( J)        2984.565

R( 3)        2985.051

R( 4)        2985.174

R( 5)        2985.295

R( 6)        2985.414

R( 7)        2985.536

R( 8)        2985.657

R( 9)        2985.777

R(10)        2985.898

******************